# Finding Interest Needle in Popularity Haystack: Improving Retrieval by Modeling Item Exposure

Rahul Agarwal
Meta Platforms, Inc.
Menlo Park, CA, USA
rahulagarwal@meta.com

Amit Jaspal
Meta Platforms, Inc.
Menlo Park, CA, USA
ajaspal@meta.com

Saurabh Gupta
Meta Platforms, Inc.
Menlo Park, CA, USA
saurabhg@meta.com

Omkar Vichare
Meta Platforms, Inc.
Menlo Park, CA, USA
ovichare@meta.com

## ABSTRACT

Recommender systems operate in closed feedback loops, where user interactions reinforce popularity bias, leading to over-recommendation of already popular items while under-exposing niche or novel content. Existing bias mitigation methods, such as Inverse Propensity Scoring (IPS) and Off-Policy Correction (OPC), primarily operate at the ranking stage or during training, lacking explicit real-time control over exposure dynamics. In this work, we introduce an exposure-aware retrieval scoring approach, which explicitly models item exposure probability and adjusts retrieval-stage ranking at inference time. Unlike prior work, this method decouples exposure effects from engagement likelihood, enabling controlled trade-offs between fairness and engagement in large-scale recommendation platforms. We validate our approach through online A/B experiments in a real-world video recommendation system, demonstrating a 25% increase in uniquely retrieved items and a 40% reduction in the dominance of over-popular content, all while maintaining overall user engagement levels. Our results establish a scalable, deployable solution for mitigating popularity bias at the retrieval stage, offering a new paradigm for bias-aware personalization.

## KEYWORDS

Recommender Systems, Popularity Bias, Item Exposure Modeling, Exposure Bias Correction, Retrieval Stage Optimization

## 1 Problem Overview

Recommender systems personalize user experiences by selecting relevant items from billions of items in milliseconds. They operate in two stages: Retrieval, optimizing recall with random negatives, and Ranking, refining precision with hard negatives [1]. These systems function as closed loops—user interactions (clicks, views, ratings) continuously refine models, shaping future recommendations.

However, this iterative feedback cycle amplifies popularity bias, where already popular items dominate recommendations, reducing diversity and limiting exposure for niche or novel content. As a result, new or less-exposed items struggle to gain visibility, leading to skewed user engagement patterns that reinforce the dominance of high-exposure content. This imbalance not only creates filter bubbles but also hinders content discovery leading to a less satisfying user experience.

## 2 Related Works

Prior approaches like Inverse Propensity Scoring (IPS) address popularity bias at the ranking stage [2, 3, 4, 5] but fail in retrieval due to severe data sparsity [4, 6]. While exposure modeling [5] and Off Policy Correction [7] have been explored in retrieval, they primarily focus on legacy collaborative filtering models and are limited to correcting bias during training. Other retrieval-stage solutions, such as negative sampling adjustments [3, 11, 12], still reinforce exposure bias by treating unseen items as irrelevant, skewing recommendations toward already-popular content.

Our work explicitly models item exposure at the retrieval stage, a first in class industrial-scale retrieval model built on top of latest model architecture and GPU accelerators [8]. Instead of passively correcting for exposure in training, we introduce an exposure-aware retrieval score that adjusts item ranking at inference, providing real-time control over popularity bias vs. engagement trade-offs. By reframing candidate selection as P(positive label | item, exposed) rather than P(positive label & exposed | item), our approach decouples engagement likelihood from exposure bias, enabling fairer recommendations while maintaining engagement. A/B tests confirm a 25% increase in uniquely retrieved items and a 40% reduction in over-recommended popular items, making this a deployable, scalable solution for real-world recommender systems.

## 3 Proposed Solution

Traditional retrieval models rank items based on P(positive label & exposure | item), inherently favoring frequently exposed items and amplifying popularity bias. To decouple relevance from exposure, we propose explicitly modeling P(exposure | item) and adjusting inference-time scoring to normalize for exposure effects, providing real-time control over popularity vs. engagement trade-offs.

Consider an example where item v1 is exposed 40 times with 20 positive interactions, while v2 is exposed 10 times with 7

positives. Standard ranking favors v1 due to higher exposure, but v2 is more likely to receive positive feedback when exposed (0.7 vs. 0.5). Our approach instead ranks based on P(positive label | item, exposed) by normalizing for P(exposure | item):

$$P(positive\ label \mid item, exposed) = \frac{P(positive\ label\ \&\ exposed \mid item)}{P\ (exposed \mid item)}$$

We train an additional exposure modeling task with randomly sampled negatives and compute the final model loss as:

$$\sum BCE(P)\ +\ w\ \cdot\ BCE(Exposed)$$

where **P** represents positive engagement signals (views, likes, shares), and **w** is the weight for the exposure task. At inference, we adjust ranking scores as:

$$score = exposure_and_positive_logit - \gamma \cdot exposure_logit$$

where subtraction in logit space corresponds to division in probability space, ensuring bias correction without impacting model efficiency.

To illustrate the end-to-end retrieval and ranking process, we employ a multi objective based retrieval model including exposure bias objective to balance trade-off between engagement, popularity bias, and relevance. As shown in Figure 1, our system integrates user, content, and contextual features into the model, which is trained on parsed user logs and candidate item distributions. The final retrieval rank is determined using a Weighted-Multiplication based scoring strategy, enabling scalable deployment in large-scale video recommendation platforms.

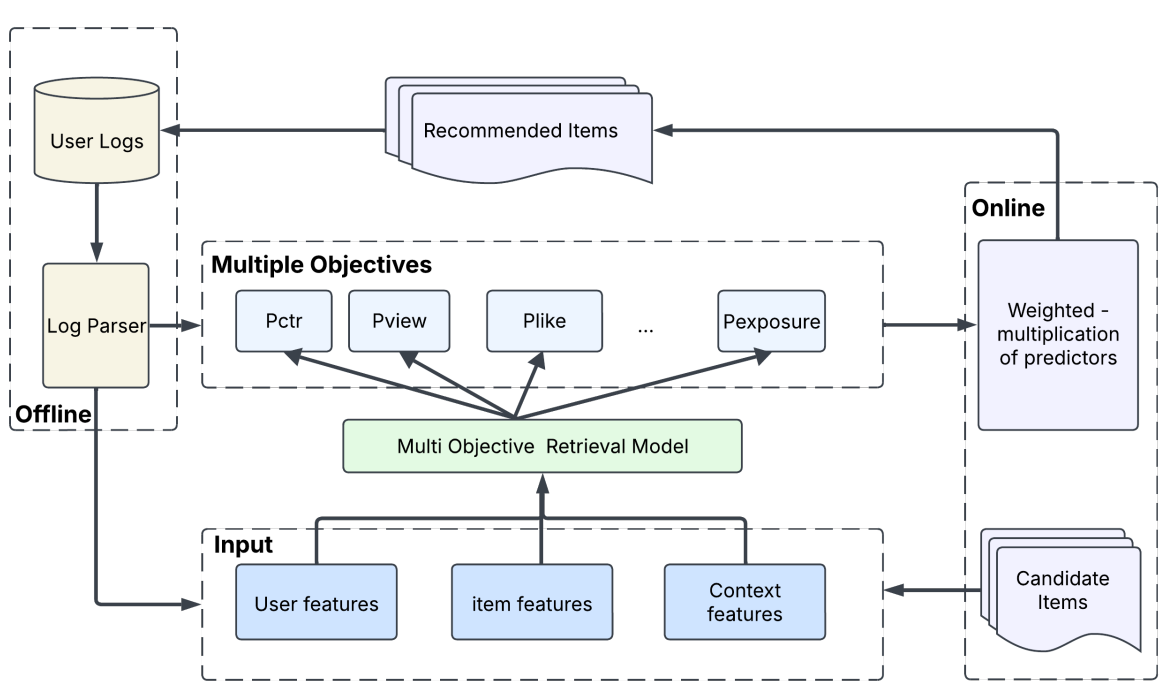

**Figure 1: End to end retrieval process**

# 4 Experiments

We conducted thorough offline and online experiments on the video recommender system to understand the impact of exposure correction on popularity bias. For offline experimentation we instrumented 2 metrics, positive recall using positive user engagements (video views, likes, shares etc.) as ground truth and negative recall using negative engagements (skipped impressions, dislikes etc.). Noted that lower values of negative recall are better. We instrumented exposure correction baseline (γ = 0.0) which ranks items purely based on observed engagement signals, leading to higher popularity bias, lower long-tail content visibility, and higher negative recall (11.0%, Table 1) due to overexposure of already-popular items. Additionally, we also instrumented single-task Off-policy correction (OPC) baseline (last column, Table 1) which applies bias correction only during training but lacks real-time exposure control at inference, resulting in the lowest positive recall (26.0%) and slightly lower negative recall (8.7%), indicating it does not effectively balance engagement and fairness. Our proposed exposure-aware scoring (γ = 0.65) achieves the best trade-off, improving positive recall (31.7%) while reducing negative recall (9.9%), making it optimal for deployment. We optimized the γ through grid search across γ ∈ [0.0, 1.0], balancing positive and negative recalls.

**Table 1: Impact of Exposure Bias Correction on Recall Metrics**

| | **γ = 0.0** | **γ = 0.65** | **γ = 1.0** | **Single task OPC baseline** |
|---|---|---|---|---|
| **Positive recall@1000** | 29.8% | 31.7% | 28.0% | 26.0% |
| **Negative recall@1000** | 11.0% | 9.9% | 7.9% | 8.7% |

While offline grid search on γ helped identify a strong candidate for online testing, we further optimized γ in a constrained space during online experiments to balance trade-offs across multiple user and creator-side business metrics before selecting the final launch value. The flexibility of our exposure-aware framework enables rapid tuning of exposure correction, making it foundational for extensive online experimentation to achieve the best balance among competing business objectives.

This approach resulted in a 25% increase in uniquely retrieved items and a 40% reduction in recommendations favoring already popular items, demonstrating its effectiveness in improving popularity bias while maintaining engagement.

# 5 Conclusion and Future Work

We introduced an exposure-aware retrieval approach that models item exposure probability, mitigating popularity bias at inference time while preserving engagement. Unlike prior ranking-stage or training-time corrections, our method enables real-time control over exposure effects, balancing fairness and engagement.

Large-scale online A/B tests in a production video recommendation system showed a 25% increase in unique item retrieval and a 40% reduction in over-recommended content, all while maintaining engagement. This demonstrates its effectiveness in enhancing niche content retrieval in large-scale industry-grade recommendation systems using GPU-accelerated inference.

While evaluated in video recommendations, our method is domain-agnostic and applicable to music, news, and e-commerce. Future work includes personalized exposure correction and longitudinal studies to refine trade-offs between fairness, engagement, and content discovery.